\documentclass[useAMS,usenatbib]{mn2e}
\usepackage{epsfig}
\title[Machine Learning for Catalogue Matching]{Applying Machine
  Learning to Catalogue Matching in Astrophysics}
\author[D J Rohde et al]{D J Rohde$^{1,2}$\thanks{E-mail:
djr@physics.uq.edu.au}, M J Drinkwater$^{1}$, M R Gallagher$^{2}$, T Downs$^{2}$ and M T Doyle$^{1}$\\
$^{1}$Department of Physics, University of Queensland, Queensland, 4072, Australia\\
$^{2}$School of ITEE, University of Queensland, Queensland, 4072, Australia\\
}

\begin{document}

\date{Accepted 1988 December 15. Received 1988 December 14; in original form 1988 October 11}

\pagerange{\pageref{firstpage}--\pageref{lastpage}} \pubyear{2002}

\maketitle

\label{firstpage}

\begin{abstract}
We present the results of applying automated machine learning techniques to the
problem of matching different object catalogues in astrophysics.  In this study
we take two partially matched catalogues where one of the two
catalogues has a large positional uncertainty.  The two catalogues we
used here were taken from the HI Parkes All Sky Survey (HIPASS), and
SuperCOSMOS optical survey.  Previous work had matched $44\%$ ($1887$
objects) of HIPASS to the SuperCOSMOS catalogue. 

A supervised learning algorithm was then applied to construct a model of the matched portion of our catalogue.  Validation of the model shows that we achieved a good classification performance ($99.12\%$ correct).  

Applying this model, to the unmatched portion of the catalogue found
$1209$ new matches.  This increases the catalogue size from $1887$ matched objects to $3096$.  The combination of these procedures yields a catalogue that is $72\%$ matched.

\end{abstract}

\begin{keywords}
catalogues, astronomical data bases : miscellaneous

\end{keywords}

\section{Introduction}

The Virtual Observatory will bring new opportunities and new
challenges.  Our study works with a problem that may become typical
in the virtual observatory context: the problem of matching catalogues with
significant positional uncertainties.

The Virtual Observatory will allow efficient access to the vast
amounts of data being collected by all sky surveys in many
wavelengths.  A fundamental operation for increasing the utility of
this data will be the matching of catalogues.  Matching catalogues will
utilise many components of the virtual observatory.  The main task of these services will be to perform a
fuzzy (probabilistic) distributed spatial join.  Distributed computing is
required so that catalogues can be published at appropriate sites all
over the world.  Special indexes have also been developed to aid in
doing fast spatial joins; at present Open SkyQuery \citep{2004adass..13..177B} is leading
progress toward making this a reality.  The study reported in this
paper is focused on the
fuzzy or probabilistic component of this problem.  That is, for a
given source, how is the correct counterpart
chosen out of a number of candidate matches within the error ellipse?
Supervised learning techniques have already been applied to the
astronomy problems of star-galaxy classification \citep{sex, 2000MNRAS.319..700A}, galaxy
morphology classification \citep{2001ApJ...548..219B} 
and the search for quasars in photometric data \citep{Richards}.  A
review paper of astronomical applications in machine learning can be found in \citet{neuralnetworks}.  Both
within astronomy and in other applications the focus of supervised
learning techniques is on \emph{regression} or \emph{pattern
  classification}.  The specific type of pattern classification problem
(the matching problem) which we consider here is reasonably novel and the
authors believe warrants further attention.  A related, but
underdeveloped field in computer science is the problem of record
\emph{linkage} \citep{fellegi}.  The solution to the problem of matching
catalogues is likely to have an impact on record linkage, which
demonstrates just one way that the development of the virtual
observatory may impact on fields outside astronomy.  Borrowing
from computer science this paper uses the term linkage to
refer to the problem of resolving the ambiguity in the matching
problem.  We draw the distinction between this problem and the
computational and network problems associated with matching
catalogues.  The database term of \emph{joining} suggests itself as being
appropriate for describing catalogue matching problems focused on the
computational or distributed nature of the problem.  In this paper we
focus on the problem of linkage.

A number of simple approaches to linkage are commonly used in
astronomy.  Often taking the closest match (in terms of position only)
is considered adequate
especially when the positional uncertainties are small, for example
\citet{1997MNRAS.284...85D} (positional uncertainties are of the order
of arcseconds).  Another more sophisticated technique, the likelihood ratio, compares the
probability that the object is a match in comparison to the
probability that it is a chance background object \citep{Sutherland}.  The
likelihood ratio itself only utilises a small number of parameters
(and only those from the more dense catalogue).  

Our work uses supervised learning techniques from machine learning in order to link
these two catalogues using all available information.  Our overall goal is
to provide a proof of concept that the full parameter list (or an
intelligently chosen subset) contains useful information that can be
used to reliably link catalogues.  While a simple method of performing
linkage using a simple supervised learning algorithm (decision trees)
has been previously demonstrated by \citet{voisin}, no follow up work on
the topic has been published.  This study offers a more complete treatment of
the problem in a number of ways.  External information is used to
construct the training set; Voisin and Donas simply used a cut on proximity to
assign labels.  We also analyse the scientific implications of different matching
algorithms and investiage different and arguably more powerful algorithms.

The linkage method we propose is well suited to a certain class of
problem.  First of all there must be a significant linkage problem; the positional uncertainty of one catalogue must be large
enough that there are frequently multiple candidate links in the more dense
catalogue.  This method also requires that there is a minimum and
maximum amount of information available.  There must be a significant
subset of the catalogues that is already linked; this is vital for us
to pursue a supervised learning procedure.  It is also important that
a significant subset of the catalogue remains unlinked in order for
the procedure to cause a significant increase in the catalogue size.

The problem we discuss here involves joining a catalogue with
comparatively poor
positional uncertainties (HICAT, \citet{meyer}) to a catalogue with good positional
uncertainties (SuperCOSMOS, \citet{hambly}).  In general the positional resolution of
the survey affects both the positional uncertainties and the density
of sources per unit area of sky.  In this study the SuperCOSMOS
catalogue is the more dense catalogue and HICAT is the sparse
catalogue.  A general statement of our problem is for each object in
the sparse catalogue to choose the correct counterpart from the dense
catalogue.  While it is not guaranteed that there is a single link in the dense catalogue, we are only dealing with the cases where we assume this to be true.

In this paper, we extend the work previously presented in \citet{rohde} by
considering the output (new matches) of the matching procedure that we
have developed.  This work is also applied to the final version of the
HOPCAT Catalogue.

This paper has the following structure.  Section 2 discusses the problem
domain that we are investigating (in particular the catalogues involved).  Section 3 discusses the
construction and validation of the model.  Section 4 discusses how we apply the model to the unmatched portion of
the HIPASS Optical Catalogue (HOPCAT) in order to match a further $1209$
objects.   Section 5 concludes by making some overall comments about our results.

\section{Problem Domain and Catalogue Details}

\subsection{HICAT}

The HI Parkes All Sky Survey (HIPASS) is a survey of the entire southern sky
for HI.  The HIPASS catalogue (HICAT) \citep{meyer} was produced by signal
processing software run over the HIPASS data cubes.  The result of this
catalogue is 4315 HI sources with accurate redshifts and significant positional
uncertainties where (RA has a $\sigma = 0.78$ arcminutes \citep{Zwaan2}).  HICAT describes
each source using many parameters, the most important of these are
velocity, peak flux ($S_p$), integrated flux ($S_{int}$) and velocity width.

\subsection{SuperCOSMOS}

SuperCOSMOS is a survey of the entire southern sky on
photographic plates taken by the UK Schmidt Telescope.  This is imaging data
and as such has accurate positions but no redshift.  

A catalogue has been produced of the SuperCOSMOS Images, the
description of the image processing used to extract this catalogue is
described in \citet{hambly2}.  For this application it was decided that it was
best to reprocess the images using the SExtractor package \citep{sex} to obtain better segmentation.  The
SuperCOSMOS parameters are area, semi-major axis, semi-minor axis, $B_j$
(mag), $R$ (mag) and $I$ (mag).  This catalogue contained a large number of stars
which obviously were non-matching, for this reason it was decided to
also provide a star-galaxy classification.  SExtractor can only
provide star-galaxy separation using its built in neural network when
the images are from a CCD rather than a photographic plate.  For this
reason the following two step procedure was used to obtain classes.
Diffraction spikes were observed as an obvious feature 
to assist in star-galaxy classification.  Software was written
using the cfitsio library which analysed the images and measured the
length of the spikes of all objects.  A training set was then
constructed of $1000$ galaxies and $1000$ stars and a support vector
machine(see Section 3.3) was
trained to classify these objects using all of the previously
mentioned SuperCOSMOS features as well as the diffraction spike
feature.  The use of machine learning techniques have been common
place for the problem of star-galaxy classification for some time
\citep{neuralnetworks, sex}.  Using a cross validation
methodology where the algorithm is tested on data that it was not
trained on the star-galaxy classifier was able to show a performance
of 88\%.

\subsection{HOPCAT}

A complementary study by \citet{doyle} produced the HOPCAT catalogue,
which matched 1887 of the 4315 HICAT sources.  The procedure for
matching involved joining the optical candidates to redshift
observations taken from the Six Degree Field survey (6dF) \citep{2003aprm.conf...97W} and the NASA
Extragalactic Database (NED).  If all of the optical candidates had redshift
information and if there was exactly one object matching the HICAT
redshift then it was deemed to be a match.  Please see \citep{doyle}
for more details.

This procedure allowed the matching of many HICAT sources to optical
counterparts.  It is however a slow procedure requiring heavy human
intervention and it also was inconclusive in cases where there was no
additional redshift information from 6dF or NED.

\section{Construction and Validation of Model}
While supervised learning algorithms automate much of the model
construction process, human judgment must be used at a number of
steps.  The choice of input variables must be made for the algorithm,
this procedure is known as feature selection.  There is no correct
procedure for doing this, except to call upon human judgment.
Learning algorithm performance is generally improved by the choice
of a small but informative set of features.

Closely related to feature selection is the preprocessing of input
variables.  For example is it advantageous to provide raw magnitudes
or colour index information?  If the distribution of a variable is not
uniform it may be advantageous to transform it prior to learning.

In this study a number of algorithms will be attempted and a procedure
known as 10-fold cross validation is used to estimate the
generalisation performance of these
models (see Section 3.3).  The best of these models is selected and performance is reported.

\subsection{Feature Selection}
In order for an optical parameter to be useful it must convey some
useful information, either a relation between the optical parameter
and radio parameters or something that will identify that the object
is a galaxy likely to be a strong HI source.  In contrast a radio parameter is only useful if it can be used to
identify a relation between radio and optical parameters.  The asymmetry above is due to the fact that there are many optical
candidate matches for a single radio object.  

In machine learning the parameters that are
selected to build a model are called features.  The rationale for
our choice of features is as follows: log area and $B_j$ (mag) should be roughly
correlated with log peak flux ($S_p$) and log integrated flux ($S_{int}$).  Velocity is also a measure of distance
so an inversely proportional relationship would be expected between
velocity and either area or magnitude.  We would expect highly
elliptical optical objects to link with radio objects with high velocity
widths. It was unclear if galaxy colour would contribute to the
classifier, although it may be a means of detecting late-type galaxies that are
likely to contain significant amounts of HI.

The only parameter not mentioned is separation, this is obviously
useful as we would expect objects with low separation to be more
likely to be matches.

The logarithm was taken of integrated flux, peak flux and area so that these
would all roughly correlate with magnitude.  A list of all the features selected as machine learning inputs is given in Table 1.

\begin{table}
\centering
\begin{tabular}{|c|c|c|}
\hline
Feature&Origin&Name\\
\hline
1&Radio-Optical&Separation\\
2&Radio&Velocity\\
3&Radio&Velocity Width\\
4&Radio&Log Integrated Flux ($S_{int}$)\\
5&Radio&Log Peak Flux ($S_p$)\\
6&Optical&Log Isophotal Area\\
7&Optical&Semi-major Axis\\
8&Optical&Semi-minor Axis\\
9&Optical&$B_j$ (Magnitude)\\
10&Optical&$B_j-R$\\
11&Optical&$B_j-I$\\
12&Optical&Star-Galaxy classification\\
\hline
\end{tabular}
\label{feat}
\caption{Selected Features for machine learning inputs}
\label{features}
\end{table}

\subsection{Framing the Matching Problem as a Pattern Classification Problem}

The matching problem is not framed automatically as a pattern classification
problem.  In order to make it one we combine inputs of radio and
optical objects into a single vector.  If the pair of objects are
matching then the vector gets a positive label, otherwise the pair is
given a negative label.  The negative training points are determined
by taking all the non-matching objects from the dense catalogue and
pairing them with the respective object in the sparse catalogue.  We
also employ a \emph{mismatched} set of negative examples which is
discussed later.

It is normal to report the error on both the negative and the positive
parts of the training set separately.  This is particularly helpful in situations
where the amount of positive and negative training data is
unbalanced (we have $6.3$ negative examples for every positive).  If a
classifier was to always give a negative response it would trivially
give a classification of $\frac{6.3}{7.3}$ or $86\%$ over all examples:
$0\%$ on the positive data and 100\% on the negative.  For this reason
the performance on the positive and negative data is reported
separately.  In order to avoid the inclusion of massive numbers of
small and faint objects, only objects with an area greater than $600$
pixels were included in this study.

Here we report success rates rather than error rates.  Error rates
over the negative data are known as false positives and error rates
over the positive data are known as false negatives.  False positives
and false negatives are related to traditional measures of
completeness and efficiency.  Both completeness and false negatives
refer to the objects that are lost from the sample due to
misclassification.  Likewise efficiency and false positives refer to
the incorrect objects that are found in our sample.

In this situation the relationships between completeness and false
negatives, and efficiency and false positives are complicated by the
framing of the problem in terms of binary pattern classification.  In
this situation the classifier is not constrained to give exactly one
match; the `combinatorial' nature of the output causes there to be no
direct relationship between completeness and false negatives and efficiency and false positives.

\subsection{Model Selection}

There are a number of supervised learning algorithms that are
appropriate to apply to this problem.  One is the Support Vector
Machine. The Support Vector Machine (SVM) computes a nonlinear mapping that
transforms its input data into a high dimensional feature space where
patterns of different classes can be separated by a hyperplane
\citep{vapnik}.  The software being used is SVM Light \citep{joachims};
this software is free for scientific use.  SVMs have a number of
parameters that can be tuned for optimal performance, including the
kernel function.  Kernel functions map the data to a high
dimensional feature space.  The SVM searches for a function that is
linear in this high dimensional space, but non-linear in input space
to separate these two classes.  Popular kernels include linear, polynomial and
radial basis functions (RBF) \citep{learningwithkernels}.

SVMs also allow the soft margin \citep{cristianini} to be adjusted which is a parameter
that controls the trade off between smooth and overly complex
functions.  Controlling this trade off is necessary to obtain good
generalization.  Functions that represent the training data well but
do not generalise to novel examples are said to have \emph{overfit} the data
in machine learning terminology.  The soft margin is a tool for the
SVM to avoid overfitting.

Another popular and older algorithm is the neural network \citep{bishop}.  Neural
networks are functions with a network-like topology and many free
parameters.  A gradient descent optimisation algorithm is used to
partially search the parameter space for a suitable representation of
the data.  There are a countless number of heuristics for improving or
altering the performance of neural networks, however in this study we
implement the simplest of these algorithms i.e. backpropagation.  The neural network is
used with $3$, $4$, $5$ and $6$ hidden units.  A neural
network without any hidden units (the perceptron) is also used.

There is no way to know a priori which algorithm will give the best
performance.  The recommended procedure is to run a battery of tests
using a good selection of candidate algorithms and parameters and
measure the generalisation ability of each.  An effective method for
getting an accurate measure of generalisation ability is the 10-fold
test.  This involves dividing the training data into 10 equal parts,
an algorithms is then trained on 9 of the subsets and tested on the
10th.  This procedure is repeated 10 times in order to average this
result over the entire dataset.  The model which gave the best
generalisation should then be selected.  This procedure is known as
cross validation.  Table ~\ref{features} shows the
generalisation performance of multiple learning algorithms with
different parameters.  The SVM has different kernels (linear,
polynomial and RBF) and different soft margins ($0.1$, $1$ and $10$).  The
neural networks have different number of hidden units (free parameters). Each
network was trained for $1000$ iterations (``epochs'').  

\begin{table*}

 \begin{minipage}{100mm}
  \caption{Performance of Algorithms and parameters}
\begin{tabular}{|l|l|l|l|l|l|} \hline
\label{diffkernels}
  Algorithm & Soft Margin & Pos Data  & Neg Data&Overall\\ 
  (Kernel)  & (c) / hu  & \% correct& \% correct&\% correct\\ \hline \hline
  SVM&0.1&87.47 $\pm$2.24&98.70 $\pm$0.27&97.17\\
  Linear&1&88.94 $\pm$2.43&98.75 $\pm$0.50&97.41\\
  &10&88.80 $\pm$2.44&98.80 $\pm$0.25&97.43\\
  \hline
  SVM&0.1&90.04 $\pm$0.31&99.04 $\pm$1.67&97.93\\
  Poly&1&94.18 $\pm$1.91&99.44 $\pm$0.20&98.72\\
  d=2&10&96.02 $\pm$1.47&99.53 $\pm$0.29&99.05\\
  \hline
  SVM&0.1&94.91 $\pm$1.93&99.46 $\pm$0.27&98.84\\
  Poly&1&96.24 $\pm$1.83&99.54 $\pm$0.20&99.09\\
  d=3&10&96.69 $\pm$1.26&99.50 $\pm$0.42&99.12 *\\
  \hline
  SVM&0.1&89.39 $\pm$2.58&99.21 $\pm$0.27&97.87\\
  RBF&1&93.66 $\pm$2.47&99.50 $\pm$0.28&98.70\\
  $\gamma=1$&10&95.43 $\pm$1.69&99.66 $\pm$0.17&99.08\\
  \hline
  Perceptron&&86.81 $\pm$7.73&97.52 $\pm$2.78&96.05\\
  \hline
  Neural Net&hu=3&93.81 $\pm$1.68&95.50 $\pm$1.41&95.27\\
  &hu=4&94.10 $\pm$2.07&95.46 $\pm$1.38&95.27\\
  &hu=5&93.50 $\pm$3.59&95.48 $\pm$1.26&95.21\\
  &hu=6&93.45 $\pm$2.01&95.62 $\pm$1.23&95.32\\
  \hline
\end{tabular}
\end{minipage}

Note:  The errors reported are the standard deviation on the
performance rate found when doing a 10-fold test.  The asterisk (*)
denotes the model with the best overall performance.  The overall
result takes in to account that there is approximately $6.3$ times as
much negative data as positive, this results in more importance being
required on classifying negative data correctly.  The overall
percentage correct is given by the formula : $R_{overall} = 0.1365
\times R_{pos}  + 0.8635 \times R_{neg}$

\medskip 

In the second column, hu refers to hidden units in a neural network.

\end{table*}

\subsection{Model Performance}
Running a battery of different algorithms showed that a SVM with a
third degree polynomial and a soft margin of 10 was optimal (see Table
~\ref{diffkernels}).  The percentages reported here are the result of
10-fold tests reported separately over positive and negative
examples.  This is useful, because sometimes the performance over the
dataset and the performance over the positives and negatives vary
considerably.  Performance (in terms of percentage correct) over the positive and negative examples are
reported separately.  As the correctly classified examples (true positives and true negatives) are in all cases distributed
over positives and negative data we can say that in all cases non-trivial models are found.

\subsection{Feature Importance}
Feature importance is the determination of how much information is
given by each input.  Feature importance is notoriously difficult \citep{guyon}.
The reason for this is that in different combinations features have
different effects, so in reality importance is a combinatorial
problem.  In the case of a matching problem the
combinatorial nature is emphasised because inputs are of interest in
the amount that they correlate with other inputs.

The measuring of the importance of the input is also highly tied to
the problem of estimating the classification model.  Adding features
can make the estimation more difficult due to the curse of
dimensionality \footnote{The curse of dimensionality refers to the
  exponential increase of hypervolume as a function of dimension.
  Finding good models (discriminating functions) that lie in a high dimensional space, is known
  to be more difficult than finding models in a lower dimensional
  space.}.  This has the potential to make the addition of useful features reduce overall classification performance.

The construction of our learning problem leads to some unusual
characteristics.  The positive learning vectors consist of variables
from the sparse catalogue and the dense catalogue joined together.
This means that the information from the sparse catalogue is repeated
for every entry candidate match in the dense catalogue.  Simulations
on input importance have shown that optical parameters alone are often
sufficient to achieve moderate classification.  In this case we obtain a classification of 94\%.  At the outset of this
project it was hoped that there would exist relationships between the
radio and optical parameters which would aid in classification.
This simulation shows that apriori, rejection of objects (stars and
galaxies) from the dense catalogue is a more powerful element of this problem.

A special mismatched dataset was introduced to test the hypothesis
that radio data could contribute \emph{any} useful information.  The dataset
consisted of the normal positive matches, plus a random sample of
radio sources matched to distant optical sources.  The separation
feature was removed from this simulation.  Without radio information
on this dataset 47\% classification was achieved, while when radio information
was added classification improved to 72\%.  This confirmed that
relationships do exist between the radio and optical parameters of
these galaxies. 

\section{Application of the Model to Unmatched Data}
\begin{figure}
  \centerline{\hbox{ \hspace{0.20in} 
    \epsfxsize=1.0in
    \epsffile{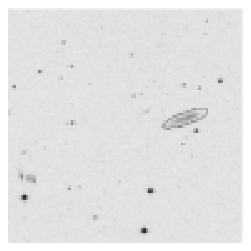}
    \epsfxsize=1.0in
    \epsffile{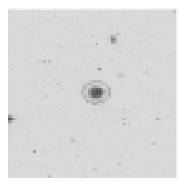}
    \epsfxsize=1.0in
    \epsffile{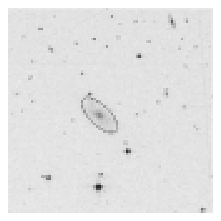}
    }
  }

  \vspace{9pt}
  \hbox{\hspace{0.2in} (a) \hspace{0.8in} (b) \hspace{0.8in} (c)} 
  \vspace{9pt}

  \centerline{\hbox{ \hspace{0.20in}
    \epsfxsize=1.0in
    \epsffile{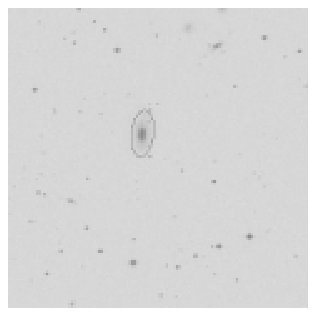}
    \epsfxsize=1.0in
    \epsffile{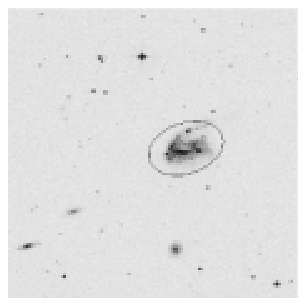}
    \epsfxsize=1.0in
    \epsffile{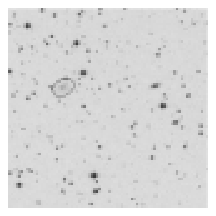}
    }
  }

  \vspace{9pt}
  \hbox{\hspace{0.2in} (d) \hspace{0.8in} (e) \hspace{0.8in} (f)} 
  \vspace{9pt}

 \centerline{\hbox{ \hspace{0.20in}
    \epsfxsize=1.0in
    \epsffile{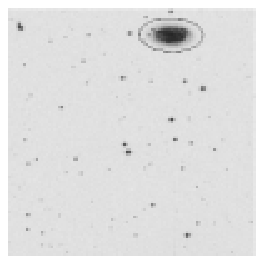}
    \epsfxsize=1.0in
    \epsffile{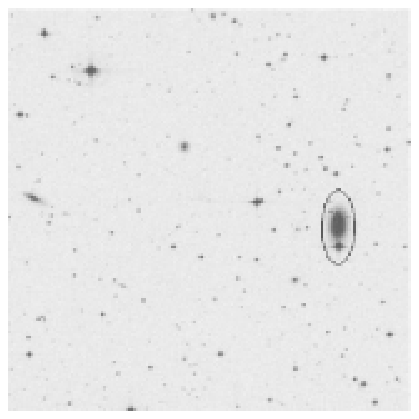}
    \epsfxsize=1.0in
    \epsffile{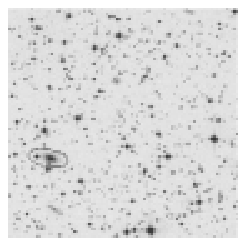}
    }
  }

  \vspace{9pt}
  \hbox{\hspace{0.2in} (g) \hspace{0.8in} (h) \hspace{0.8in} (i)} 
  \vspace{9pt}

  \caption{A sample of the new matching objects}
  \label{newobjects}
\end{figure}

\begin{figure}
  \centerline{\hbox{ \hspace{0.20in} 
    \epsfxsize=1.0in
    \epsffile{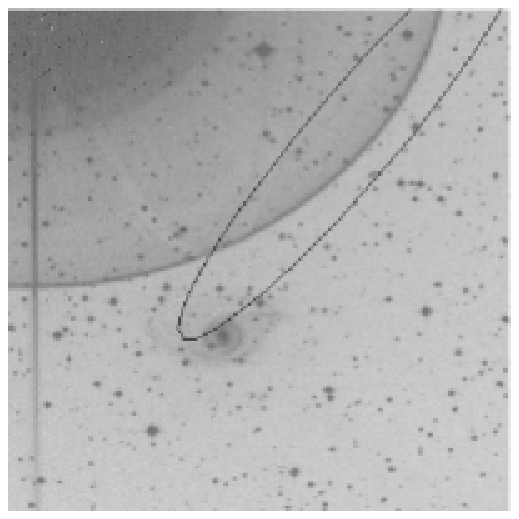}
    \epsfxsize=1.0in
    \epsffile{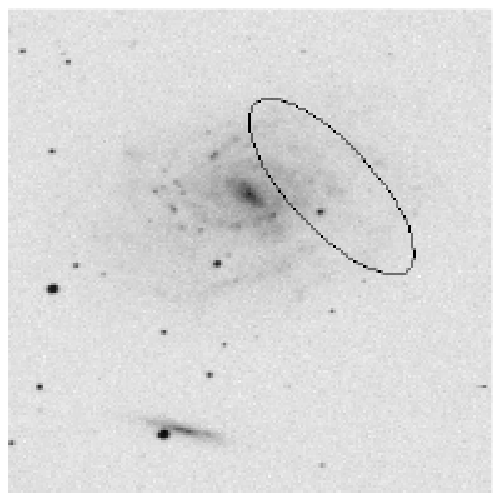}
    \epsfxsize=1.0in
    \epsffile{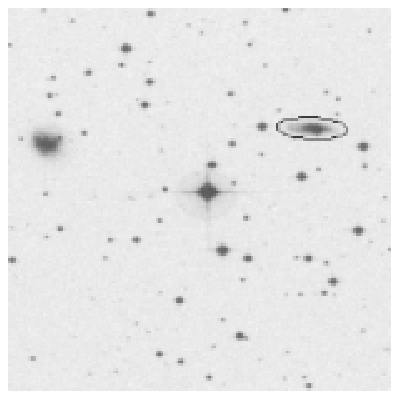}
    }
  }

  \vspace{9pt}
  \hbox{\hspace{0.2in} (a) \hspace{0.8in} (b) \hspace{0.8in} (c)} 
  \vspace{9pt}

  \centerline{\hbox{ \hspace{0.20in}
    \epsfxsize=1.0in
    \epsffile{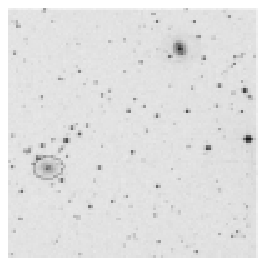}
    \epsfxsize=1.0in
    \epsffile{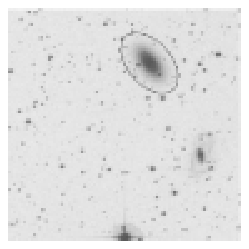}
    \epsfxsize=1.0in
    \epsffile{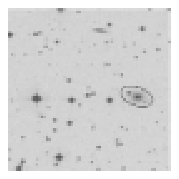}
    }
  }

  \vspace{9pt}
  \hbox{\hspace{0.2in} (d) \hspace{0.8in} (e) \hspace{0.8in} (f)} 
  \vspace{9pt}

 \centerline{\hbox{ \hspace{0.20in}
    \epsfxsize=1.0in
    \epsffile{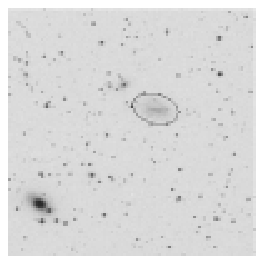}
    \epsfxsize=1.0in
    \epsffile{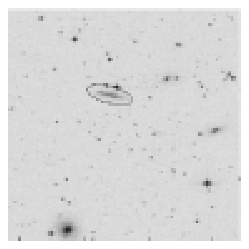}
    \epsfxsize=1.0in
    \epsffile{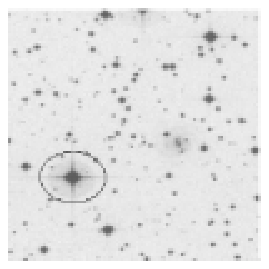}
    }
  }

  \vspace{9pt}
  \hbox{\hspace{0.2in} (g) \hspace{0.8in} (h) \hspace{0.8in} (i)} 
  \vspace{9pt}

  \caption{ Misclassified objects. a), b) are spurious ellipses marked as matches.  c) d) and e) optical measurements of the velocities of the galaxies show that machine learning chose the wrong galaxy.  f) g) and h) machine learning disagrees with HOPCAT, but there is insufficient information to establish which is correct.  i) a bright star at close proximity is chosen as a match.}
  \label{miscobjects}
\end{figure}

In order to apply our binary classification model to a HICAT source it
must be evaluated against every candidate match in the HICAT region of
uncertainty.  This increases the chance of error from the above
estimate because many model evaluations are required.  There is also
the chance that the classifier will find no matches, a single match or
multiple matches.  The performance measures given previously were for
binary classification problems.  The statistics of false positives and
false negatives are highly related to, but are not measures of completeness and efficiency.  

In order to see
how well our model applies to the actual problem we examine only the
unique velocity matches and test what agreement level this has with
the HOPCAT catalogue.  We take only the $1608$ unique velocity matches,
out of $1887$ (this has an immediate bearing on the completeness of the
catalogue).  A sample of images of newly matching objects is shown in
Fig ~\ref{newobjects}.  Of the 1608 only 9 are misclassified, indicating that
the catalogue has high efficiency.

Accurate estimates of completeness and efficiency are not possible in this
case for three reasons.  The training data, and the data to which we
apply the model have slightly different distributions.  Our
classifier output is a binary output over each output, allowing for
ambiguous situations such as multiple matches to exist.  A cross
validation method (taking in to account unique matches) should be
applied to produce this estimate.  Finally we do not know how accurate the labels
on our training data are (HOPCAT).

By ignoring cases of multiple matches we are able to sacrifice
completeness for efficiency.  We only get a false match when there are
exactly two matches (one false positive and one false negative).  This
provides a level of error checking and means that the classifier is
not applied to the difficult or ambiguous examples.

It is noteworthy that an error here requires exactly two errors over all the candidates
and one of these must be on the match.  The 9 misclassified objects are shown in Fig ~\ref{miscobjects}.  This provides a form of error
checking as if there are multiple matches then the chance that either
the classifier has failed or that the match is ambiguous is high.  A sample of images correctly classified are shown in Fig ~\ref{newobjects} and the 9 images incorrectly classified are shown in Fig ~\ref{miscobjects}.

HOPCAT contained 2221 objects which had insufficient information to
match.  It is this data that we wish to extract new information from,
by matching it using machine learning.

The machine learning model found $1209$ of these were assigned
unique matches by the model.  The high accuracy on the test
set suggests that a very high proportion of these matches are
correct.  

\begin{figure}
\includegraphics[width=3.8in]{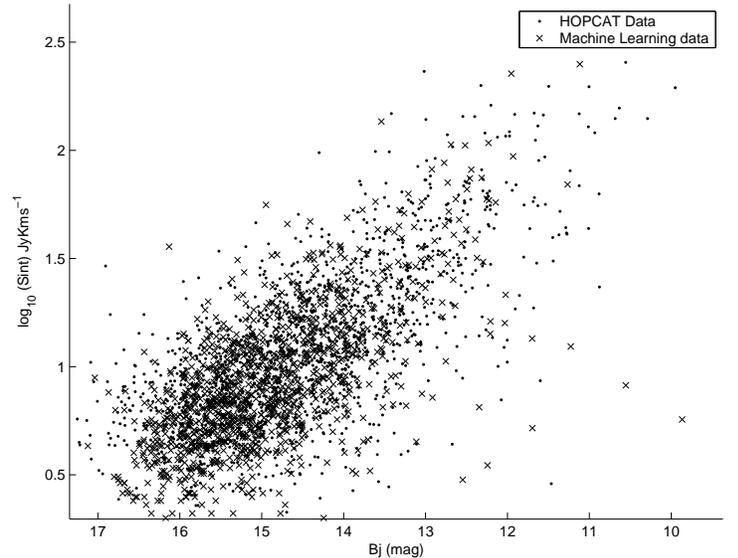}
\caption{Magnitude Flux plot of old and newly obtained datapoints}
\label{newdata}
\end{figure}

A plot of radio flux against optical
magnitude of the old and new points is shown in Fig ~\ref{newdata}.  The new data points appear to follow the same trend as the old
data points.  Although it is obvious that the two distributions are
different: the new points are more likely to be fainter in both the
optical and radio flux.  This is most likely due to a selection effect
where the training data contains brighter objects.  It appears that
the model is successfully extrapolating to fainter objects than the
training data.  The authors would like to stress that the quality of
the machine learning model should be judged on the cross-validation
performance, not the good agreement found here.

\begin{figure}
\center
\includegraphics[width=3.4in]{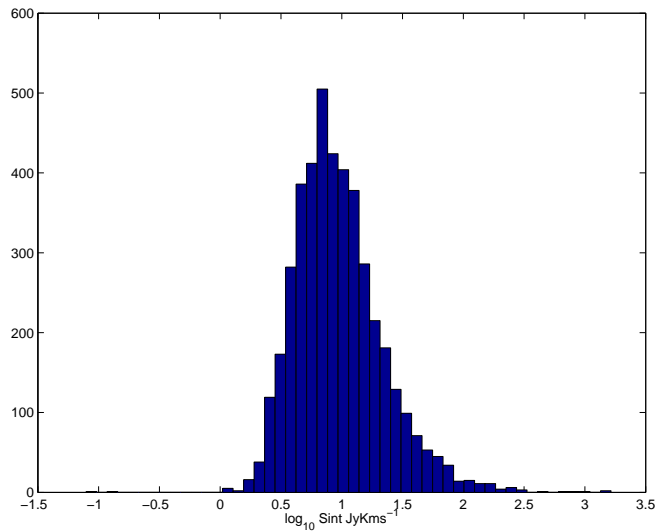}
\caption{Distribution of integrated flux ($S_{int}$)}
\label{sinthist}
\end{figure}

\begin{figure}
\center
\includegraphics[width=3.4in]{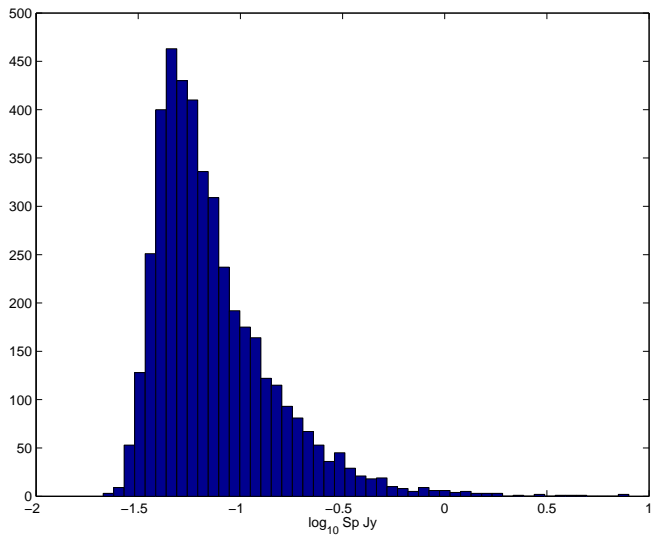}
\caption{Distribution of peak flux ($S_p$)}
\label{sphist}
\end{figure}

The SuperCOSMOS catalogue goes deeper than HICAT.  The non-linear
detection limits on HICAT can be seen in the distribution of
integrated flux ($S_{int}$) Fig ~\ref{sinthist} and peak flux ($S_p$) Fig
~\ref{sphist}.  The effect of this threshold is that objects with
an $S_{int}$ $ < 0.5 JyKms^{-1}$ are under represented in Fig ~\ref{newdata}, while
the limit on optical magnitude is low enough to have negligible
effect.  This may be responsible for a subtle curve upwards for the
faint end of the spectrum in Fig 3.

Over the $216$ blank fields $6 (3\%)$ had one or more match on them.  This
gives a rough indication of the frequency of false positives.

\section{Discussion}
The matching of catalogues can be framed as a supervised learning,
pattern classification problem.  Despite differences between matching
and pattern classification the algorithms performed remarkably well on
this data, showing performance over $99\%$.  The model we found produced
the most
discriminating power from the optical (dense) catalogue, however we
were able to show that important relations existed between the two catalogues.

This method was successful in generating $1209$ new matches to the
HOPCAT Catalogue, bringing the total number of matches to $3096$ out of
$4315$.  For a significant portion of the HICAT sources it is difficult
or impossible to find a match because there are many optical
counterparts; or the optical counterparts are obscured by the zone of avoidance.

The quality of both the source of the training data (HOPCAT) and the additional counterparts found using
machine learning, need to be verified using high resolution radio data
from the Australian Telescope Compact Array.  Verification of some or
all of the data would further validate the methods used here.

This work uncovers a number of new avenues to investigate further.
There are simple methods that could be applied to get a probability
that each candidate is a match.  This would allow assumptions such as
allowing at most one match to be built in to the classifier.

The selection effects that could be caused by such a method are
potentially complex.  The newly matched data-points are likely to show
similarity to points in the training data.  This opens up two
questions.  Firstly, if we do not have any rare objects in the training
data, then we are probably unlikely to find these objects in the newly
matched data.  Moreover if our new data points resemble our old
datapoints, what aspects of the new distribution of points are simply
resemblance to the old data, and what aspects are giving us new
information, not in the original sample?

\section{Acknowledgments}
This research has been supported by a University of Queensland
Research Development Grant and an Australian Research Council Linkage
Infrastructure Equipment and Facilities Grant.

This research has made use of the NASA/IPAC Extragalactic Database (NED) which is operated by the Jet Propulsion Laboratory, California Institute of Technology, under contract with the National Aeronautics and Space Administration.

The authors would like to thank the HIPASS Multibeam Group for
access to an early release of HIPASS and for assistance throughout the project.

\bibliographystyle{mn2e}
\bibliography{mybib}

\end{document}